\documentclass[12pt]{article}
\usepackage{wasysym, xcolor}
\usepackage{epsfig}
\usepackage{amssymb}
\newsavebox{\ns}
\newsavebox{\dbrane}
\newsavebox{\dbshort}

\def\appendix{{\newpage\section*{Appendix}}\let\appendix\section%
        {\setcounter{section}{0}
        \gdef\thesection{\Alph{section}}}\section}

\newcommand\ba{\begin{eqnarray}}
\newcommand\ea{\end{eqnarray}}

\def\Dslash{\,\,{\raise.15ex\hbox{/}\mkern-12mu D}}
\def\Dbarslash{\,\,{\raise.15ex\hbox{/}\mkern-12mu {\bar D}}}
\def\delslash{\,\,{\raise.15ex\hbox{/}\mkern-9mu \partial}}
\def\delbarslash{\,\,{\raise.15ex\hbox{/}\mkern-9mu {\bar\partial}}}
\def\pslash{\,\,{\raise.15ex\hbox{/}\mkern-9mu p}}
\def\calDslash{\,\,{\raise.15ex\hbox{/}\mkern-12mu {\cal D}}}

\def\frac#1#2{{{{#1}}\over{{#2}}}}
\newcommand{\hh}{{1\over 2}}

\renewcommand{\ll}{_}
\newcommand{\uu}{^}
\newcommand{\pp}{\partial}

\renewcommand{\exp}[1]{\, {\rm exp}\{#1\}}

\renewcommand{\d}{\delta}

\newcommand{\s}{\sigma}
\renewcommand{\t}{\tau}
\newcommand{\G}{\Gamma}

\newcommand{\e}{\epsilon}

\newcommand{\sqd}{^2}

\renewcommand{\hh}{{1\over 2}}

\newcommand{\eee}[1]{\ba{#1}\ea}

\renewcommand{\t}{\tau}

\renewcommand{\b}{\beta}

\newcommand{\D}{\Delta}

\newcommand{\pr}{^\prime {}}

\newcommand{\apr}{{\alpha^\prime} {}}

\newcommand{\IZ}{\relax\ifmmode\mathchoice
{\hbox{\cmss Z\kern-.4em Z}}{\hbox{\cmss Z\kern-.4em Z}}
{\lower.9pt\hbox{\cmsss Z\kern-.4em Z}} {\lower1.2pt\hbox{\cmsss
Z\kern-.4em Z}}\else{\cmss Z\kern-.4em Z}\fi} \font\cmss=cmss10
\font\cmsss=cmss10 at 7pt
\newcommand{\inbar}{\,\vrule height1.5ex width.4pt depth0pt}
\newcommand{\IC}{{\relax\hbox{$\inbar\kern-.3em{\rm C}$}}}
\newcommand{\IQ}{{\relax\hbox{$\inbar\kern-.3em{\rm Q}$}}}
\newcommand{\IP}{\relax{\rm I\kern-.18em P}}

\newcommand{\ed}{\dot{e}}

\newcommand{\cc}{{\cal C}}

\renewcommand{\cc}{{c_1}}

\renewcommand{\cc}{c}

\renewcommand{\pr}{{}^\prime{}}

\newcommand{\IR}{\relax{\rm I\kern-.18em R}}
\def\blfootnote{\xdef\@thefnmark{}\@footnotetext}

\renewcommand{\cc}[1]{\cite{#1}}

\newcommand{\bm}{\begin{matrix}}
\renewcommand{\em}{\end{matrix}}

\newcommand{\lno}{\left .}

\newcommand{\rba}{\right |}

\newcommand{\rr}[1]{(\ref{{#1}})}
\newcommand{\bbb}{\ba\begin{array}{c}}
\renewcommand{\eee}{\nonumber\end{array}\ea}
\newcommand{\een}[1]{\label{#1}\end{array}\ea}

\newcommand{\prpr}{^{\prime\prime}{}}

\def\hilo{{}_{{}_{{}_{{}_{{}_{}}}}} {}^{{}^{{}^{}}}}

\newcommand{\heading}[1]{\begin{center}\it {#1} \rm \end{center}}

\def\lrdd{\left ( ~}
\def\rrdd{\hilo \right )}

\newcommand{\kket}[1]{\left | {#1} \right \rangle }

\def\bi{\begin{itemize}}
\def\ei{\end{itemize}}

\def\ed{\end{document}}

\def\cc{{\cal C}}

\renewcommand{\rr}[1]{(\ref{#1})}

\def\cc{\,}

\def\xxn{\\ \\}

\def\cc{\,}

\definecolor{Red}{rgb}{1,0,0}
\newcommand{\nts}[1]{}
\newcommand{\edA}[1]{{#1}}
\newcommand{\remA}[1]{}
\newcommand{\subA}[2]{{#1}}
\begin{document}

\begin{titlepage}
\begin{flushright}
IPMU-10-0104\\
\end{flushright}
\vspace{15 mm}
\begin{center}
  {\Large \bf  Bounds for State Degeneracies  \\ in \\
 \vspace{2mm}  2D Conformal Field Theory
}
\end{center}
\vspace{6 mm}
\begin{center}
{ Simeon Hellerman and Cornelius Schmidt-Colinet}\\ ~ \\
\vspace{6mm}
{\it   Institute for the Physics and Mathematics of the Universe\\ Todai Institutes for
Advanced Study (TODIAS) \\
The University of Tokyo \\
 Kashiwa, Chiba  277-8582, Japan\\}

\end{center}
\vspace{6 mm}
\begin{center}
{\large Abstract}
\end{center}
\noindent

In this note we explore the application of modular invariance
in 2-dimensional CFT to derive universal bounds for quantities
describing certain state degeneracies,
such as the thermodynamic entropy, or the number of marginal operators.
We show that the entropy at inverse temperature $2\pi$ satisfies a universal
\it lower \rm bound, and we enumerate the principal obstacles to deriving
\it upper \rm
bounds on entropies or quantum mechanical degeneracies for
completely general CFT.  We then restrict our attention to infrared-stable
CFT with moderately low central charge, in addition to the usual assumptions of
modular invariance, unitarity and discrete operator spectrum.
For CFT in the range $c\ll{\rm L} + c\ll{\rm R} < 48 $ with no relevant operators,
we are able to prove an upper bound on the
thermodynamic entropy at inverse temperature $2\pi$.
Under the same conditions we also prove that a CFT can have no
more than $\lrdd
{{c\ll{\rm L} + c\ll{\rm R}}\over{48 - c\ll{\rm L} - c\ll{\rm R}}}
\rrdd\cdot \exp{{+ 4\pi}} - 2$ marginal deformations.
\vspace{1cm}
\begin{flushleft}
\today
\end{flushleft}
\end{titlepage}
\tableofcontents
\newpage

\section{Introduction}

Conformal field theory plays several distinct roles
in our understanding of nature, including quantum gravity.
First, 2D CFT defines the classical solutions of
the mysterious system whose perturbative amplitudes are given by the
dynamics of relativistic strings.  Second of all, conformal field theory
defines the Hamiltonian of quantum gravity in spaces with
negative cosmological constant and asymptotically anti-de Sitter
boundary conditions.  Third of all, conformal field theory
describes fixed points of the renormalization group flow that
organizes the behavior of all local quantum systems, including string theory in the
low-energy approximation.  Beyond these important roles, CFT
appears in many other guises in various aspects of fundamental physics.

It is therefore of interest to understand as clearly as possible the
gross features of the ``landscape'' of conformal field theories
\cite{douglasqfts, landscape}.
Even the case that is by far best explored, that of two-dimensional
CFT, is far from being understood in a systematic way.

Recently multiple
independent
lines of development have converged to probe the consistency conditions
on the space of conformal field theories in extreme limits.  In particular,
various papers over recent years have \subA{investigated}{probed} the question of how big a
gap can ever be opened in the spectrum of operator dimensions, above
a universal sector defined by products of stress tensors.

This question is explored under various simplifying assumptions in different
articles: in
\cite{hohnone,hohntwo,threedqgadscftwitten}, the two-dimensional case is
explored under the assumption of holomorphic factorization of the Hilbert
space; in \cite{mooreetal} the two-dimensional case is explored again
and extended superconformal symmetry is assumed; in
\cite{rattazzi, opdims} the authors examine the maximum possible dimension of
the lowest-dimension operator appearing in the operator product expansion of
two scalar operators whose dimensions are taken as given; in \cite{prev}
the two-dimensional case is probed once again and the dimension of the
lowest primary operator is bounded in a completely general CFT satisfying
the minimal properties of unitarity and discrete operator spectrum.

In all cases the result is qualitatively the same: if one fixes
the input parameters
 -- in two dimensions, the central charge, and in the case of \cite{rattazzi, opdims}
the dimensions of the external scalar operators in the OPE -- the
lowest-dimension operator that appears can never have a dimension higher
than some universal bound.\footnote{In \cite{polchinski} the converse is shown: By making
the central charge sufficiently large, it is possible to push the gap
to infinity and obtain a consistent limit where the theory contains only
a sector of low-dimension operators.}    Strengthening the assumptions -- adding
special conditions such as supersymmetry or holomorphic factorization --
improves the bound numerically but does not qualitatively change the
result.

In each case discussed above, the constraining principle is either modular
invariance, or else associativity of the OPE, which is closely related.  Both
are consistency conditions expressing the constraint that the theory must
make sense when quantized in two different, inequivalent channels -- that is,
foliating the space with two different time-slicings whose leaves may
be orthogonal to one another.  It would appear that the condition of
covariance among channels, or democracy among foliations, is a fundamental
principle that makes it impossible to pick and choose the spectrum of a CFT
at will.  In particular, the condition of modular invariance, or channel
covariance of the OPE, is incompatible 
with an attempt to deform the spectrum in an extreme way.

In all cases discussed above, the gap in operator dimensions is the
quantity under examination,
that is bounded by the principle of modular invariance, or
channel covariance more generally.
There are other ``extreme directions''
in the space of possible spectra, in which we suspect it may be
impossible to engineer the spectrum of a consistent CFT.  In particular,
one expects that a CFT of fixed central charge ought not to have
a number of marginal deformations greater than some universal
number depending on the central charge.  That is, for a 
CFT of central charge $c\ll{\rm tot}$ and discrete spectrum, it would appear likely that
there may be a fundamental limit on the dimension of moduli space.

This idea is correlated with commonly held beliefs
in mathematics and physics.  In physics, the holographic principle of 't Hooft
and Susskind \cite{holo1, holo2, bousso}
could be vitiated by limits in which the number of massless
species is pushed infinitely large~\cite{species, bousso}.
In mathematics, it is thought that
there is likely an upper bound on the Euler number of compact Calabi-Yau
threefolds \cite{gross, hunt, chlo}, which would follow immediately
from a fully general
bound on the number of marginal operators of a superconformal field
theory of central charge $c\ll{\rm L} = c\ll{\rm R} =
c = {3\over 2} \hat{c} = 9$.

In this note we derive rigorous bounds on certain state degeneracies:
First, a lower bound on the thermodynamic entropy at
the inverse temperature $\b= 2\pi$ that maps to itself under the
modular $S$-transformation.
Second, we prove an upper bound on
the number of
marginal operators of a two-dimensional CFT of a given central
charge, under certain conditions.  Under the same conditions
we also derive an upper bound on the
thermodynamic entropy at inverse temperature $\b  = 2\pi$.

The bounds proven herein,
though requiring certain conditions to hold that are less than
fully general, provide an illustration of a principle
bounding state degeneracies from above, that the authors hope may
apply in a broader set of circumstances.

\section{Energy and entropy bounds from modular invariance}

\subsection{Review of modular invariance}

The principle underlying our bounds is modular invariance, in
particular invariance under the modular $S$-transformation
$\t\to - {1\over\t}$ of
the partition function of the CFT on a two-torus with complex
structure $\t$.  The partition function of a two-dimensional
CFT on such a torus can be written
as
\bbb
Z[\t] = {\rm tr} \lrdd \exp{ 2\pi i \t (L\ll 0 - {{c\ll {\rm R}}\over{24}}
) - 2\pi i \bar{\tau} (\tilde{L}\ll 0 - {{c\ll{\rm L}}\over{24}} )} \rrdd \ ,
\een{hampartfunc}
where $L\ll 0$ and $\tilde{L}\ll 0$ are the zeroth right- and left-moving
Virasoro generators, $c\ll {\rm R}$ and $c\ll{\rm L}$ are
the right- and left-moving central charges, 
and the complex structure
$\tau$ lies in the upper half plane.  The torus can be thought of as
a quotient of the complex plane $\IC = \{ \s\equiv \s\uu 1 + i \s\uu 2\}$
by the identifications $\s\sim \s + 2\pi \sim \s + 2\pi \tau$.  The
generators $L\ll 0$ and $\tilde{L}\ll 0$ are related to energy $H$ and
momentum $P\ll 1$ by
\bbb
L\ll 0 = \hh (H + P\ll 1) + {{c\ll{\rm R}}\over {24}}\ ,
\xxn
\tilde{L}\ll 0 = \hh (H - P\ll 1)+ {{c\ll{\rm L}}\over {24}}\ .
\eee
We can represent this partition function as a path integral on a torus
with metric
\bbb
\left ( \begin{matrix} { g\ll{11} & g\ll{12} \cr g\ll{12} & g\ll{22} }
\end{matrix} \right ) =  {1 \over{{\rm Im}~\t}} \left ( \begin{matrix} { 1  &  {\rm Re}~\t
\cr {\rm Re} ~\t & |\t|\sqd } \end{matrix} \right )
\eee
Modular invariance of the partition function
is the statement that a local conformal field
theory has nothing to distinguish the various cycles of the two-torus
other than the background metric itself.  Therefore, partition
functions on two distinct background metrics differing
by a large coordinate transformation should have the same
partition function.  A large coordinate transformation acts on the
cycles of the torus as
\bbb
SL(2,\IZ)\ni \G = \left [ \begin{matrix} { a & b \cr c & d} \end{matrix}
\right ]\ ,
\eee
which induces an action $\t\mapsto {{a \t + b }\over{c\t + d}}$ on the
complex structure of the torus.

The modular group is generated by two elements, $S \equiv \left [
\begin{matrix} { 0 & -1 \cr 1 & 0 } \end{matrix} \right ]$ and
$T \equiv \left [ \begin{matrix} { 1 & 1 \cr 0 & 1 } \end{matrix} \right ]$,
which satisfy the relation $(ST)\uu 3 = -1$.  For modular invariance to
be a good symmetry of the CFT at the quantum level, it suffices to
check that the partition function transforms under both $S$ and
$T$ without anomalous phases.

For purposes of this note, as in \cite{prev}, we shall not require
invariance under the $T$-transformation, but only under the $S$-transformation.
Failure of the partition function to be invariant under the
$T$-transformation is easily understood in the Hamiltonian framework as
a failure of the momentum $P\ll 1$ to be quantized in integer units;
the expression \rr{hampartfunc} is invariant under $\t\to \t + 1$
if and only if every state in the spectrum has $P\ll 1 \in \IZ$.

The $S$-transformation is both more obscure and more robust.

It is
obscure in the sense that it is completely non-manifest in the
Hamiltonian formalism.  That is to say, a Hamiltonian formulation of
a quantum theory depends on a foliation of spacetime,
with Hilbert spaces
associated to the individual leaves and Hamiltonian flow implementing
linear transformations between the Hilbert spaces on different leaves.  But the
$S$-transformation is not a canonical transformation, and
does not preserve even a single leaf of any smooth foliation
of the torus, even up to Hamiltonian flow.  The $S$-transformation, even when an
exact symmetry of the system, is \it not \rm realized as an
action on the Hilbert space of quantum states.

On the other hand, the $S$-transformation is a good
symmetry of the quantum theory under 
very broad conditions: if the theory
has a Poincar\'e-invariant path integral formulation in terms of local variables, then the
path integral automatically respects the $S$-transformation at the quantum level.
There are many known theories that satisfy all axioms of CFT except symmetry
under the $S$-transformation; however all such examples involve
imposing projections directly on a Hilbert space and do not
have partition functions defined by a path integral over local variables.
In this paper we shall restrict our attention to modular-invariant
CFT, though only invariance under the $S$-transformation is necessary for
our results to hold.

\subsection{Review of previous work}

\heading{The medium-temperature expansion}

Going forward let us assume that the CFT is described by a unitary
quantum mechanics (\it i.e., \rm a Hermitean Hamiltonian 
with a positive definite norm on the Hilbert space), and that the spectrum
of the Hamiltonian in finite volume is discrete.  Thus the
partition function can be written as
\bbb
Z[\b] = {\rm tr}\lrdd \exp{- \b H} \rrdd = \sum\ll n \exp{- \b E\ll n} \ ,
\een{thermal}
where $E\ll n$ are the discrete, real eigenvalues of the Hamiltonian
$H$ of the theory on a circle of length $2\pi$.

By virtue of Cardy's formula, the partition function and all of its
derivatives are convergent for any positive $\b$.
The $p\uu{\rm{\underline{th}}}$ derivative is equal to a sum
of derivatives of exponentials, that is
\bbb
\lrdd {{\pp}\over{\pp\b}} \rrdd\uu p Z[\b] = (-1)\uu p\cc {\rm tr} 
 \lrdd H\uu p \exp{- \b H} \rrdd = (-1)\uu p \sum\ll n E\ll n\uu p \cc \exp{- \b E\ll n} \ .
\een{}

For  purely imaginary complex structure $\t = {{i\b}\over{2\pi}}$
expression \rr{hampartfunc} reduces to the usual
thermodynamic partition function \rr{thermal}
at inverse temperature $\b$.  Then the $S$-transformation acts
on the partition function as
\bbb
Z[\b] \to Z[{{4\pi\sqd}\over\b}]\ .
\een{}
A modular invariant partition function is invariant under this transformation:
\bbb
Z[\b] = Z[{{4\pi\sqd}\over\b}]\ .
\een{modinv}
It follows immediately \cite{prev} that
\bbb
\lno (\b\pp\ll\b)\uu p Z[\b] \rba\ll{\b = 2\pi} = 0\ .
\een{}
for any positive odd $p$.
Written in terms of the energies $E\ll n$, this can be
written
\bbb
\sum\ll{n} \exp{- 2\pi E\ll n} \cc f\ll p (E\ll n) = 0\ ,
\een{medtemp0}
where $f\ll p(E)$ is a $p\uu{\rm{\underline{th}}}$ order polynomial
defined by
\bbb
f\ll p (E) \equiv \exp{+ 2\pi E} \cc
\lno \lrdd \b{{\pp}\over{\pp\b}} \rrdd\uu p \cc \exp{- \b E}
\rba\ll{\b = 2\pi}\ .
\een{}
Explicit expressions for low $p$ are:
\bbb
f\ll 1(E) = - 2\pi E
\\ \\
f\ll 3(E) = - (2 \pi E)\uu 3 + 3 (2\pi E)\sqd - (2\pi E)
\een{explicitderived13}
More generally, if $F(x)$ is any odd function of $x$, then
we define a derived function
\bbb
f\ll F (E) \equiv \exp{+ 2\pi E} \cc
\lno F\lrdd \b {{\pp}\over{\pp\b}} \rrdd \cdot
\exp{- \b E} \rba\ll
{\b = 2\pi} \ .
\een{}
The polynomials $f\ll p$ are just the derived functions corresponding
to $F(x) = x\uu p$.

For any odd $F$ the derived function $f\ll F(E)$ satisfies
\bbb
\sum\ll n \exp{- 2\pi E \ll n} f\ll F(E\ll n) = 0\ ,
\een{medtemp1}
where $\{E\ll n\}$ is the spectrum of a modular invariant CFT.
For some purposes it is convenient to think of this condition in terms
of a density \nts{(Maybe we should change $\rho$ to $\tilde{\rho}$ because $\rho$
should really always stand for the actual spectral density, IMHO.)}
\bbb
\rho(E) \equiv \exp{- 2\pi E} \cc \sum\ll n \d(E - E\ll n)\ ,
\een{rhodef}
which is $\exp{- 2\pi E}$ times the usual spectral density.  Then we can
write the condition \rr{medtemp1} as
\bbb
\int dE\cc \rho(E) \cc f\ll F(E) = 0
\een{medtemp2}
for an $f$ derived from any odd $F$.
By virtue of Cardy's formula, the partition
function is real analytic as a function of $\b$ for any $\b$.  Therefore the conditions \rr{medtemp0}
for all odd $p$ are
not only a consequence of invariance under the $S$-transformation,
but when taken together are sufficient to imply it as well. \nts{(Does this
statement need to be rethought?  Let's be a little careful about radii of convergence
and so on.)}

The condition \rr{medtemp0} follows directly from writing
\bbb
\b \equiv 2\pi \exp{s} \ ,
\eee
noting that the $S$-transformation acts as $s \mapsto -s$, and
expanding the equation \rr{modinv} to $p\uu{\rm{\underline{th}}}$ order
in $s$.  Since we are expanding the partition function in the neighborhood
of $\b = 2\pi$, which is intermediate between the
high-temperature $\b\to 0$ and low-temperature $\b\to \infty$
r\'egimes,
we refer to the expansion in $s$ as the \it medium-temperature~\rm
expansion \cite{prev}.  The medium-temperature expansion has proven to be useful in
deriving general constraints on the spectrum of a modular invariant
conformal field theory \cite{cardy, cardy2, prev}.

\heading{Review of the upper bound on $\D\ll 1$}

In \cite{prev} we used the medium-temperature expansion to prove that any
conformal field theory satisfying unitarity and modular invariance,
with a discrete spectrum, satisfies the bound
\bbb
\D\ll 1 \leq {{ c\ll{\rm tot}}\over{12}} + 0.48\ ,
\een{}
where
\bbb
c\ll{\rm tot} \equiv c\ll{\rm L} + c\ll{\rm R}
\een{}
and $\D\ll 1$ is the dimension of the primary operator of lowest dimension
other than the identity itself.
For $c\ll{\rm tot} < 24 - {{18}\over \pi}\simeq 18.270$, the proof is
completely elementary and does not depend on any use of
representation theory of the Virasoro algebra.  We recall the proof here.

Fix a value of the central charge $c\ll{\rm tot}$ less than
$24 - {{18}\over{\pi}}$.  Now consider the cubic polynomial
\bbb
F(x) \equiv x \cc (x\sqd - 4\pi\sqd E\ll 0 \sqd + 6\pi E\ll 0 - 1)\ ,
\een{}
where $E\ll 0$ is the ground state energy
\bbb
E\ll 0 = - {{c\ll{\rm tot}}\over
{24}} .
\een{e0val}

Using expressions \rr{explicitderived13}, the derived polynomial is
\bbb
f\ll F(E) = f\ll 3 (E) - (4\pi \sqd E\ll 0 \sqd - 6\pi E\ll 0 + 1)
f\ll 1(E)
\\ \\
= - 8\pi\uu 3 \cc E \cc (E - E\ll 0)\cc (E - E\ll +)\ ,
\een{derpol}
where $E\ll + \equiv {3\over{2\pi}} - E\ll 0$.
By equation \rr{medtemp2}, the quantity
\bbb
\int \ll {E\ll 0} \uu\infty f\ll F(E) \rho (E) dE
\een{exampleident}
must vanish.
The measure $\rho(E)$ is positive and the
derived polynomial
$f\ll F(E)$ vanishes at $E\ll 0$ and is negative for $E > E\ll +$.  If
all excited energy levels were to be $E\ll +$ or higher,  then the quantity
\rr{exampleident} could not vanish, as it would receive
only negative contributions.  The lowest
excited level $E\ll 1$ must therefore be lower than $E\ll + = {3\over{2\pi}}
- E\ll 0$.  Translating into operator dimensions via
$\D = E - E\ll 0$, we find that the lowest-dimension operator
other than the identity can have dimension no higher than
$\D\ll + \equiv E\ll + - E\ll 0 = {3\over{2\pi}} - 2 E\ll 0$.  Using
\rr{e0val} we write the bound as
\bbb
\D\ll 1 <  \D\ll +  \equiv {3\over{2\pi}} + {{c\ll{\rm tot}}\over{12}}\ .
\een{firstbound}

For $c\ll{\rm tot} < 24 - {{18}\over \pi}$ the value of
$\D\ll +$ is less than $2$, and it follows that the operator
of dimension $\D\ll 1$ cannot be a descendant of the identity; it must
therefore be primary or else the descendant of a primary operator of
even lower dimension.  So in this range of central charge,
we learn that equation \rr{firstbound} can be taken to apply specifically to
the dimension of the lowest \it primary~\rm
operator above the identity.

\heading{Medium-temperature equations with characters}

  For higher central charge a similar bound
can be proven using roughly the same argument,
with the contributions to the partition function organized according to
full representations of the Virasoro algebra rather than individual
energy levels.  The proof for $c\ll{\rm tot} > 24 - {{18}\over{\pi}}$
uses some elementary facts about the representation theory
of the Virasoro algebra, in particular formulae for the characters of the
Virasoro algebra for various representations \cite{kacdet, feiginfuks, details}.

It is natural to generalize the
derived polynomials $f\ll p(E)$ to derived polynomials $f\ll{p\cc | \cc \chi}$
with respect to characters $\chi$ of the Virasoro algebra.  In this
context, we will only consider characters restricted to the imaginary
axis of their argument, $\chi(\b)$ with $\t = {{i \b}\over{2\pi}}$.

Our simplifying assumptions, that $c\ll {\rm L}, c\ll{\rm R} > 1$ and
that there are no holomorphic or antiholomorphic operators, guarantee
that there are only two types of representations of the Virasoro
algebra for a unitary CFT with discrete spectrum \cite{kacdet, details}:

\bi
\item{the conformal family of the vacuum, generated by the independent
states
\bbb
{{\prod}\atop{{m,n \geq 2}}  }
(L\ll{-m}) \uu{N\ll m} (\tilde{L}\ll{-n})\uu{\tilde{N}\ll n} \kket 0\ ,
\een{}
and
}
\item{the conformal family of a generic primary with dimension $\D$, generated
by the independent states
\bbb
{{\prod}\atop{{m,n \geq 1}}  }
(L\ll{-m}) \uu{N\ll m} (\tilde{L}\ll{-n})\uu{\tilde{N}\ll n} \kket \D\ ,
\een{}
}
\ei
with the $N\ll m, \tilde{N}\ll n$ running over
all possible $N\ll m, \tilde{N}\ll n \geq 0$ in each case.

Each type of conformal family is then spanned by different
\subA{monomials}{polynomials} in Virasoro raising operators acting on the primary state, with
each monomial raising the energy of the vacuum by an amount
$ \sum\ll n n \cc N\ll n + n \tilde{N}\ll n$.  The contributions of
the two families to the partition function are then given by
\bbb
\chi\ll v(\b) \exp{- E\ll 0 \b}
\een{}
for the vacuum family, and
\bbb
\chi\ll g (\b) \exp{- (\D + E\ll 0) \b}
\een{}
for the family of the generic primary, with the vacuum and generic characters
given by
\bbb
\chi\ll v (\b) \equiv \prod\ll{n \geq 2} (1 - \exp{- n \b})\uu{-2}\ ,
\\ \\
\chi\ll g (\b) \equiv \prod\ll{n \geq 1} (1 - \exp{- n \b})\uu{-2}\ ,
\een{}
respectively.
Then the full partition function can be written as
\bbb
Z[\b] = \chi\ll v(\b) \exp{- \b E\ll 0} + \chi\ll g(\b)(\sum\ll {n=1}
\uu\infty \exp{- \b E\ll n})\ ,
\een{}
where the sum in the second term runs over energies $E\ll n \equiv
\D\ll n + E\ll 0$ of \it primary states \rm alone.  Then invariance under
the modular $S$-transformation $\b \to 4\pi\sqd / \b$ can be expressed
in terms of energies of primary states.  For a given
function $F(x)$ and any character $\chi$, define the derived functions
\bbb
f\ll{F\cc | \cc \chi}(E)  \equiv {{\exp{+ 2\pi E}}\over{\chi(2\pi)}}  \cc \lno
F \lrdd \b\pp\ll\b \rrdd \cdot [\chi(\b) \cc \exp{- \b E}]\rba\ll{\b
= 2\pi}\ .
\een{}
Then the invariance under the modular $S$-transformation can be expressed
by the condition that
\bbb
\chi\ll v (2\pi) \exp{- 2\pi E\ll 0} f\ll {F\cc | \cc \chi\ll v}(E\ll 0)
+ \sum\ll {n = 1}\uu\infty \chi\ll g (2\pi ) \exp{- 2\pi E\ll n} \cc
f\ll{F\cc | \cc \chi\ll g}(E\ll n) = 0\ ,
\een{}
for any odd $F$.  To express this condition in terms of measures,
define
\bbb
\rho\ll v (E) \equiv \chi\ll v (2\pi) \exp{- 2\pi E} \d(E - E\ll 0)
\een{}
and
\bbb
\rho\ll g (E) \equiv \chi\ll g (2\pi) \exp{- 2\pi E} \sum\ll{n = 1}\uu
\infty \d(E - E\ll n)\ ,
\een{}
where $n$ runs over non-vacuum primary states.  Then the condition for
modular invariance is
\bbb
\int dE\cc \rho\ll v (E) f\ll {F\cc | \cc \chi\ll v}(E) + \int dE\cc
\rho\ll g (E)
f\ll {F\cc | \cc \chi\ll g}(E) = 0
\een{medtempchars}
for any odd $F$.  For an arbitrary character $\chi(\b)$, the derived
polynomials $f\ll{p\cc | \cc \chi}$
corresponding to low-order monomials
$F(x) = x\uu p$ are
\bbb
f\ll{1\cc | \cc \chi}(E) = - (2 \pi  E)  + 2\pi {{\chi\pr(2\pi)}\over{\chi(2\pi)}}
\\ \\
f\ll{3\cc | \cc \chi}(E) = - (2\pi E)\uu 3  + (2\pi E)\sqd \lrdd 6\pi {{\chi\pr(2\pi)}
\over{\chi(2\pi)}} + 3 \rrdd
\\ \\
- (2\pi E) \lrdd
12 \pi\sqd {{\chi\prpr(2\pi)}\over{\chi(2\pi)}} + 12 \pi {{\chi\pr(2\pi)}
\over{\chi(2\pi)}} + 1 \rrdd
+ \lrdd 8\pi\uu 3 {{\chi\uu{\prime\prime\prime}(2\pi)}\over{\chi(2\pi)}}
+ 12\pi\sqd {{\chi\prpr(2\pi)}\over{\chi(2\pi)}} + 2\pi{{\chi\pr(2\pi)}
\over{\chi(2\pi)}} \rrdd
\een{lowordercharpols}

In \cite{prev} we use this structure to
derive a bound on the weight of the lowest non-vacuum
primary dimension $\D\ll 1$ that applies for arbitrarily high values of
the total central charge.  We refer the reader to
\cite{prev} for details.  We want to emphasize, however, that the
generalized proof is not much more complicated than the elementary proof
for low central charge that we have reviewed in the previous subsection.

\subsection{A lower bound for the entropy at medium temperature}

The
most straightforward consequence of modular invariance is
to give a \it lower, \rm rather than an upper bound for the thermodynamic
entropy of the canonical ensemble, particularly at inverse temperature $\b = 2\pi$.

From the first order of the medium-temperature expansion,
it follows that a universal lower bound
holds
for the thermodynamic entropy at medium temperature $\b\uu{-1} = {1\over{2\pi}}$.
The entropy $\s$ is
related to the partition function via
\bbb
\s = {\rm ln}(Z) + \b \left \langle E \right \rangle\ .
\een{}
Using the derived
function $f\ll 1(E) = - 2\pi E$ in equation \rr{medtemp0}, we have
\bbb
\lno \left\langle E \right\rangle\rba \ll{\b = 2\pi} = 0\ .
\een{}
By unitarity, every contribution to the partition function is positive,
so the value of $Z$ is bounded below by its vacuum contribution
$\exp{- 2\pi E\ll 0}$, so we have
\bbb
\lno \s\rba \ll{\b = 2\pi} \geq - 2\pi E\ll 0  =  {{\pi c\ll{\rm tot}}\over{12}} \ .
\een{finalentropylower}

We have given an elementary proof of a universal lower bound for the thermodynamic entropy in a modular invariant 2D CFT at a particular temperature.
On the other hand we shall see in the next section
that there can be no fully general upper bound on
the thermodynamic entropy or on the microstate degeneracies that would hold
without imposing additional
assumptions on the CFT.  Understanding the issues involved will help
us to formulate a useful set of additional assumptions on the CFT
as we go forward.

\newcommand{\hedd}[1]{\heading{#1}}
\section{Meta-problem: Why are upper bounds for state degeneracies hard?}

In the previous section we reviewed the derivation of
an upper bound on the energy of
the first excited energy level.  Should it not be possible, then, to
prove an upper bound on entropy -- thermodynamic entropy or quantum mechanical
degeneracies -- using similar techniques?  Let us examine, briefly, a
few reasons why the type of argument in the previous section
cannot be generalized very easily to give a fully general upper bound for
entropy or quantum mechanical degeneracies, without imposing additional
assumptions on the CFT as inputs.

\hedd{ (a) The homogeneity problem.}

The homogeneity problem is
a meta-problem with any candidate for a method to bound the
entropy above, using invariance under the $S$-transformation.
Suppose we had some equation of type \rr{medtemp2} that would \it
always be
violated \rm if the entropy -- the thermodynamic entropy or the
quantum mechanical
degeneracy, in some energy range,
according to whatever definition -- were to be
sufficiently high.  We can argue by contradiction that no such equation can
ever exist.

Suppose such an equation did exist, that ruled out
the possibility of a modular invariant spectrum with effective state degeneracy
greater than $n\ll{\rm max}$, by
whatever definition.  But we could always take $k$ copies of a modular
invariant spectrum with effective state degeneracy $n < n\ll{\rm max}$,
such that $k\cc n > n\ll{\rm max}$.  Taking 
$k$ copies of a modular invariant
spectrum automatically yields a modular invariant spectrum, so it is clear
that there can be no direct constraint from modular invariance bounding
the effective state degeneracy above, without using additional inputs.

\hedd{(b) The continuum problem at the vacuum.}

One way around the homogeneity problem is to use the fact that
the spectrum under consideration is not arbitrary, but corresponds
to the spectrum of a good CFT, satisfying all the usual CFT axioms
including cluster decomposition.  Together with unitarity and the
state-operator correspondence,
cluster decomposition implies
that there is a unique lowest state, with
energy $E\ll 0 = - {{c\ll{\rm tot}}\over{24}}$.

Then it may be
possible in principle to find an upper bound on the entropy (or the
degeneracy of some states in some range of energy) that evades the
homogeneity problem by using the fact
that the degeneracy of the vacuum is never greater than 1.

To exploit this fact, it would be necessary to find an odd function $F$
such that the derived function $f\ll F(E)$ gets a positive contribution
from the vacuum, a negative contribution from the states of interest, and
a sum of contributions from all other states that is bounded
above independently.  (In particular, it would not do to pick an $F$
such that $f\ll F (E\ll 0) = 0$, as we did to prove the universal upper bound
on $E\ll 1$.)

This type of approach to bounding the entropy
suffers from a separate meta-problem that we shall call the
\it continuum problem. \rm  By the continuum problem, we mean that
such a proof could never apply in cases where the spectrum
develops an approximate continuum of states with energies an arbitrarily
small amount above the vacuum.  In such
a case the continuum would contribute with the same sign as the vacuum
(by continuity of $f(E)$)
with an unboundedly large coefficient, due to the presence of an arbitrarily
numerous set of levels in the range between $E\ll 0$ and $E\ll 0 + \e$.

Of course, no CFT under our consideration ever has a strict
continuum; we are always assuming
that our CFT have a discrete operator spectrum, or equivalently discrete
spectrum of the Hamiltonian in finite volume.  But many CFT are known
to come in families with singular limits where the limiting
spectrum has a continuum of some kind.  In particular, many
familiar moduli spaces of CFT have limits in which the CFT
can be thought of as a sigma
model with volume approaching infinity.  So the continuum problem
is a general no-go principle for upper bounds on
CFT degeneracies of any sort that do not use additional consistency
conditions of the CFT, or assume a minimum gap in the
spectrum above the vacuum.

\hedd{(c) The hyperfine structure problem of character corrections.}

To evade the continuum problem without assuming a minimum gap in the
spectrum as an input, we could try using other consistency conditions
of the CFT; in particular, we could try using the organization of the
spectrum into representations of the Virasoro algebra.  However
this approach immediately runs into the problem that the
differences between different characters of the Virasoro algebra
are numerically very tiny.

As in
\cite{prev}, simplify the discussion by assuming that
$c\ll {\rm L}$ and $c\ll {\rm R}$ are both greater than 1, and
that there is no chiral algebra of the theory other than the Virasoro
algebra.  The condition \rr{medtempchars} then treats
the vacuum conformal family, which contributes to $\rho\ll v$, differently
from the conformal families of the other primaries, which contribute
to $\rho\ll g$.  Therefore it is possible in principle to find
odd $F$ such that the derived polynomials $f\ll {F\cc | \cc \chi\ll v}(E\ll 0)$
and $f\ll{F\cc | \cc \chi\ll g}(E)$ contribute with different sign, even
when $E$ is arbitrarily close to $E\ll 0$.  In practice, however, it
seems difficult to generate a useful bound this way.

An approach to overcoming the continuum problem based on the difference
in Virasoro representations between that of the vacuum and that
of the generic primaries lying in the continuum just above the vacuum,
would need to exploit the differences between the derived
functions $f\ll{F\cc | \cc \chi\ll v}$ and $f\ll{F\cc | \cc \chi\ll g}$, evaluated
at $E\ll 0$.  But
the absolute difference between the derived functions for the two
different types of conformal family is numerically small,
being proportional to $\exp{- 2\pi} \simeq {1\over{535}}$.

For instance, suppose we want to take the lowest-order polynomial
possible, $F(x) = x\uu 1$.  Then the two derived polynomials are
\bbb
f\ll{1\cc | \cc \chi\ll v } (E) =  - 2\pi E + 4\pi \sum\ll{n = 2}\uu \infty
{{n\cc \exp{- 2\pi n}}\over{1 - \exp{- 2\pi n}}}
\xxn
f\ll {1\cc | \cc \chi\ll g}(E) =
- 2\pi E + 4\pi \sum\ll{n = 1}\uu \infty
{{n\cc \exp{- 2\pi n}}\over{1 - \exp{- 2\pi n}}}
\een{}
The two differ only by
\bbb
f\ll{1\cc | \cc \chi\ll g}(E) - f\ll{1\cc | \cc \chi\ll v}(E) = 4\pi{{\exp{- 2\pi}}\over
{1 - \exp{- 2\pi}}} \simeq 2.35 \times 10\uu{-2} \ .
\een{}
Furthermore,
a minimal criterion for overcoming the continuum problem is that
the value of ${{\rm lim}\atop{E\to E\ll 0}} f\ll{F\cc | \cc \chi\ll g}(E) $
should have a different sign from $f\ll {F\cc | \cc \chi\ll v}(E\ll 0)$.
In the case $F(x) = x$, it is impossible for the two to
differ in sign unless $E\ll 0 > 0$, which would be inconsistent with
positivity of the central charge, and thus with unitarity.

\hedd{(d) The fine structure problem at large central charge}

There is a separate problem in attempting to bound the entropy from
above by using invariance of the partition
function under the modular $S$-transformation, that is particularly
acute when the central charge becomes large.  To see the nature
of the problem, consider in particular an attempt to bound the degeneracy
of marginal operators in the limit of large central charge.
When the central charge
is large, it becomes increasingly difficult to find derived functions
$f\ll  F(E)$ that are positive for $E = E\ll 0$, and negative for
all $E \geq E\ll 0 + 2$.  In the limit $E\ll 0 \to - \infty$
with $E / E\ll 0$ held fixed, the derived function $f\ll F(E)$ is to good
approximation equal to $F(-2\pi E)$, which is an \it odd \rm function.
For $F(x) = x\uu p$, we have
\bbb
f\ll{p}(E) = (- 2\pi E)\uu p  + O(E\uu{p-1}).
\een{}
Thus if we take an odd polynomial $F(x)$ such that
$f\ll F(E)$ is positive at $E = E\ll 0$ but negative at $E\ll 0 + 2$,
then it will tend to be positive again by the time it reaches
$E = -(E\ll 0 + 2)$, if $|E\ll 0|$ is large.

The even part of the function $f\ll F(E)$ is a kind of ``fine structure''
of subleading order at large central charge.  But it is only by
tuning the form of the function $F(x)$ to enhance the contribution of
the even part of $f\ll F$ relative to the odd part,
that one has any hope of obtaining a derived function
with properties that could imply a bound.

\hedd{(e) The resolutional problem. }

The \it resolutional problem \rm
is the difficulty in formulating a suitable definition
of the entropy of marginal operators that
is robust against small perturbations of the spectrum.
Certainly it seems quite unlikely
that there should be any unitary conformal field theory, with
discrete spectrum and
central charges $c\ll L = c\ll R = 2$, say, with $10\uu{10\uu{100}}$
marginal operators, for instance.  And yet, if we alter the question slightly
to ask whether there might be a unitary conformal field theory
with discrete spectrum and
central charges $c\ll L = c\ll R = 2$, and $10\uu{10\uu{100}}$
operators between dimensions $2 - \e$ and $2 + \e$, for arbitrarily
small values of $\e$, the answer is that, \it yes, \rm there
certainly do exist such CFT.  Simply consider a sigma model with
target space $T\uu 2$, with cycles of length $2\pi R$, and the
kinetic term for the target space coordinates $X\uu a$ normalized as
${\cal L} = {1\over{4\pi\apr}} g\uu{ij} (\pp i X\uu a) (\pp j X\uu a)$.

The theory
contains scalar primary operators of the form ${\cal O}\ll{n\ll 1, n\ll 2 }
\equiv :\exp{i n\ll a X\ll a / R }:$, which have
dimension $\D = {{\apr}\over{2 R\sqd}}n\ll a \sqd$.
For $R$ much larger than $\sqrt{\apr}$, the set of operators with
$ 2 - \e \leq \D \leq 2 + \e$ corresponds to the set of integer
pairs $(n\ll 1,n\ll 2)$ lying in the Euclidean plane between two
spheres centered at the origin, with radii $\sqrt{{{2 R\sqd (2 - \e)}}\over
{\apr}}$ and $\sqrt{{{2 R\sqd (2 + \e)}}\over
{\apr}}$.  The region of interest is an annular region
of radius $2 {R\over{\sqrt{\apr}}}$, circumference
$4 \pi { R  \over{\sqrt{\apr}}}$ and thickness ${{ R\e}\over{\apr}}$.
The lattice points are distributed with unit density in the plane, so
the annular region contains of order ${{4\pi R\sqd \e}\over{\apr}}$
lattice points.  Thus, no matter how small $\e$ is chosen to be,
the radius $R$ can always be made sufficiently large that the number
of almost-marginal operators -- scalar operators with dimensions between
$2 - \e$ and $2 + \e$ can be made as large as desired.

\vskip.2in

The five problems described above
are not entirely logically independent from one another.
The strict version of problem $(a)$ can be solved trivially by
assuming cluster decomposition, but even then this solution is ``unstable''
against turning into problem $(b)$ under a small perturbation of
the spectrum.  Problem $(b)$ can in principle be
solved by organizing the partition function
using characters of the Virasoro algebra to separate the vacuum from the
continuum.  But in practice one runs into problem (c), that
the effect of the distinct
characters at $\b = 2\pi$ is so small that it is not easy to exploit the
separateness of these contributions to derive a bound.

Problems
(a)-(c) all concern the difficulty of controlling unwanted contributions
to \rr{medtemp0}-\rr{medtemp2} from energy levels near the vacuum $E\ll 0$.
Problem (d) concerns the difficulty of controlling unwanted contributions
from energy levels much higher than the marginal operators
at $E\ll 0 + 2$, and problem
(e) concerns the difficulty of controlling unwanted contributions
from levels arbitrarily close to $E\ll 0 + 2$.


The purpose of discussing these meta-problems is not only
to warn ourselves away from too-naive attempts to prove a bound using
modular invariance, but also to guide ourselves in choosing a
favorable set of additional assumptions on the CFT that will
allow us to avoid these persistent difficulties.

\section{Upper bounds for state degeneracies under certain conditions}

In this section we will
prove a bound on the number of marginal operators in an
infrared stable CFT with $c\ll{\rm L} + c\ll{\rm R}$ less than 48.
We will also derive an upper bound on the thermodynamic entropy at inverse temperature
$2\pi$, under the same conditions.
We conclude with a discussion of the possibility of bounding the number
of marginal operators under other sets of assumptions, such as
extended superconformal symmetry.

\subsection{A bound on the number of marginal operators}

Keeping in mind the meta problems discussed in the section above, we will
now choose some simple assumptions that will allow us to avoid
problems (a) through (e).
We can avoid problems (a)-(c) by assuming \edA{cluster decomposition, and} the
absence of a continuum of states just above $E\ll 0$:
in fact the bound is simplest
if we restrict our considerations to infrared stable fixed points
of the renormalization group -- CFT with no relevant operators
at all other than the identity.  We will avoid problem (d) by restricting
our considerations to CFT with moderately low central charge -- say,
less than 24 \remA{or so }on the left and on the right, so
$c\ll{\rm tot} < 48$.  We need not make
any additional assumption in order to evade problem (e) -- in fact we will
see that the assumptions of infrared stability and $c\ll{\rm tot} < 48$
are enough to prove a bound; problem (e) is avoided automatically.

Under these assumptions it is possible to prove a bound using
only the first order in the medium-temperature expansion.  Let $N$
be the number of primary operators of dimension 2.  The degeneracy at
energy $E\ll 0 + 2$ is then $N+2$, with the 2 extra operators coming
from the left- and right-moving stress tensor.  The number $N$ includes
both scalar primaries of weight $(1,1)$
as well as spin-$1$ operators of weights $(3/2,1/2)$
and $(1/2, 3/2)$.  The simple method we are using is not sufficiently
refined to distinguish these, but an upper bound on $N$ necessarily bounds
the number of scalar primary operators, so it will not matter too much
that we do not distinguish them.

 Taking
$F(x) = - (x/(2 \pi))$ gives $f\ll F (E)=   E$, and we have the
equation
\bbb
0 =  E\ll 0 \exp{- 2\pi E\ll 0}
\xxn
 + (E\ll 0 + 2) (N + 2)
\exp{- 2\pi (E\ll 0 + 2)}
\xxn
+ \sum
(E\ll 0 + \D) \exp{- 2\pi (E\ll 0 + \D)}
\een{maineq}
where the sum in the third term runs over all operators with dimension
$\D > 2$.
For $c\ll{\rm tot} < 48$, the quantity $E\ll 0 + 2 = {{48 - c\ll{\rm tot}}
\over{24}}$ is positive, so the only negative contribution in
equation \rr{maineq} is
the first.  Multiplying \rr{maineq} through by $\exp{+ 2\pi(E\ll 0 + 2)}$
we obtain
\bbb
0 < (E\ll 0 + 2) (N + 2)
\xxn
< (E\ll 0 + 2) (N + 2)
+ \sum
(E\ll 0 + \D) \exp{- 2 \pi(\D - 2)}
\xxn
 = - E\ll 0 \exp{+ 4\pi }\ ,
\een{}
giving us a bound
\bbb
N <
\lrdd
{{c\ll{\rm L} + c\ll{\rm R}}\over{48 - c\ll{\rm L} - c\ll{\rm R}}}
\rrdd\cdot \exp{{+ 4\pi}} - 2\ .
\een{}
This result depends on the assumption that
there are no operators with dimensions $\D$ between $0$ and $2$.
Due to the earlier result \cite{prev}, reviewed in the second section,
this can only be the case if $c\ll{\rm tot} > 24 - {{18}\over{\pi}}
\simeq 18.27$.  So the interesting range of central charge, where
a useful bound may be proven from modular invariance
at first order in the medium-temperature expansion, is
\bbb
\sim 18.27 < c\ll{\rm tot} < 48\ .
\een{}
For integer values of $c\ll{\rm tot}$ in this range, we include
a table giving the maximum possible number of marginal
operators in a CFT with no relevant operators:

\def\comma{\hskip -.05in ,\cc}
\begin{table}[ht]
\begin{minipage}[b]{0.5\linewidth}
\centering
\begin{tabular}{| c || c |}
\hline
$c\ll{\rm tot}$ & $N\uu{\rm max}$ \\
\hline\hline
 19 & 187 \comma 869 \\
 20 & 204 \comma 820 \\
 21 & 223 \comma 026 \\
 22 & 242 \comma 633 \\
 23 & 263 \comma 809 \\
 24 & 286 \comma 749 \\
 25 & 311 \comma 684 \\
 26 & 338 \comma 885 \\
 27 & 368 \comma 678 \\
 28 & 401 \comma 449 \\
 29 & 437 \comma 671 \\
 30 & 477 \comma 916 \\
 31 & 522 \comma 897 \\
 32 & 573 \comma 500 \\
 33 & 630 \comma 850 \\
\hline
\end{tabular}
\end{minipage}
\hspace{0.5cm}
\begin{minipage}[b]{0.5\linewidth}\centering
\begin{tabular}{| c || c |}
\hline
$c\ll{\rm tot}$ & $N\uu{\rm max}$ \\
\hline\hline
 34 & 696 \comma 394 \\
 35 & 772 \comma 020 \\
 36 & 860 \comma 251 \\
 37 & 964 \comma 525 \\
 38 & 1 \comma 089 \comma 652 \\
 39 & 1 \comma 242 \comma 587 \\
 40 & 1 \comma 433 \comma 754 \\
 41 & 1 \comma 679 \comma 541 \\
 42 & 2 \comma 007 \comma 257 \\
 43 & 2 \comma 466 \comma 059 \\
 44 & 3 \comma 154 \comma 262 \\
 45 & 4 \comma 301 \comma 267 \\
 46 & 6 \comma  595 \comma 278 \\
 47 & 13 \comma \cc 477 \comma 309 \\
 48 &   $\infty$  \\
\hline
\end{tabular}
\end{minipage}
\end{table}\label{tableone}
As an immediate corollary of our bound, note that the resolutional
problem, problem (e) of the previous section, has resolved itself
automatically: through modular invariance as expressed at first
order in the medium-temperature expansion,
the assumption of no relevant operators other than the identity 
implies the absence of a continuum of states
with dimensions near 2.
If the latter did exist, then equation
\rr{maineq} would receive an unboundedly large number of
positive contributions
going as $ (E\ll 0 + 2 + \e) \cc \exp{- 2\pi(E\ll 0 + 2 + \e)}$,
for $\e$ made arbitrarily small, without offsetting negative contributions;
if there were a (near)-continuum close to $\D = 2$ but no states between
$\D = 0$ and $\D = 2$, then the first-order medium-temperature condition
for modular invariance could not be
satisfied.
\subsection{An upper bound on the entropy at medium temperature}

To complement our discussion of a lower bound for
thermodynamic entropy in the second section, we would
like to establish an upper bound on the thermodynamic entropy of the canonical
ensemble at medium temperature, $\b = 2\pi$, under certain conditions.
We impose the same assumptions on our CFT as we did to derive the upper
bound on the degeneracy of marginal operators in the previous subsection:
That is, in addition to unitarity, cluster decomposition,
discrete operator spectrum and invariance
under the modular $S$-transformation, we also assume that $c\ll{\rm tot} < 48$
and that there are no relevant operators other than the identity.

As in the previous subsection, we need consider only the leading nontrivial order
in the medium-temperature expansion in order to derive an interesting bound.
Using the fact that $E\ll 0 < 0$, we can write $E\ll 0 = - |E\ll 0|$, and
equation \rr{maineq} can be written as
\bbb
1 = \sum\ll{n \geq 1} {{E\ll n}\over{|E\ll 0|}}\cc \exp{- 2\pi(E\ll n + |E\ll 0|)}\ .
\een{nextmaineq}
We are assuming $0 < c\ll{\rm tot} < 48$, and that $2 \leq \D\ll 1 < \D\ll 2 < \cdots $.
Then using $E\ll n = \D\ll n + E\ll 0 = \D\ll n - {{c\ll{\rm tot}}\over{24}}$, we have
\bbb
0 \leq E\ll 1 \leq E\ll 2 \cdots
\een{}
So the $n\uu{\rm{\underline{th}}}$ term on the
right hand side of \rr{nextmaineq} is no smaller than
$ {{E\ll 1}\over{|E\ll 0|}} \cc \exp{- 2\pi (E\ll n + |E\ll 0|)}$.  Summing terms,
we find
\bbb
{{E\ll 1}\over{|E\ll 0|}} \cc \sum\ll{n\geq 1} \exp{- 2\pi (E\ll n + |E\ll 0|)}
\leq \sum\ll{n \geq 1} {{E\ll n}\over{|E\ll 0|}}\cc \exp{- 2\pi(E\ll n + |E\ll 0|)} = 1\ .
\een{}
Multiply each side by $ {{|E\ll 0|}\over {E\ll 1}}\cc \exp{+ 2\pi |E\ll 0|}$ to derive
the inequality
\bbb
\sum\ll{n \geq 1} \exp{- 2\pi E\ll n} \leq {{|E\ll 0 |}\over{E\ll 1}}
\cc  \exp{+ 2\pi |E\ll 0|}\ .
\een{}
We may have no information about the specific value of $E\ll 1$, but under our
assumptions we do know
that $E\ll 1 \geq 2 -  |E\ll 0| > 0$, so
\bbb
\sum\ll{n \geq 1} \exp{- 2\pi E\ll n} \leq {{|E\ll 0|}\over{2 - |E\ll 0|}}\cc
\exp{+ 2\pi |E\ll 0|}\ .
\een{}
Adding the vacuum contribution $\exp{- 2\pi E\ll 0 } = \exp{+ 2\pi |E\ll 0|}$ yields
an upper bound for the full partition function at $\b = 2\pi$:
\bbb
Z[2\pi]
= \sum\ll{n\geq 0} \exp{- 2\pi E\ll n}
\xxn
\leq \lrdd 1 + {{|E\ll 0|}\over{2 - |E\ll 0|}}
\rrdd \cc\exp{+ 2\pi |E\ll 0|} = {{48}\over{48 - {c\ll {\rm tot}}}} \cc\exp{ + {{\pi c\ll{\rm tot}}\over{12}}}\ .
\een{penulti}
The bound \rr{penulti} on the partition function
in turn
gives us an upper bound for the thermodynamic entropy of the canonical ensemble at
medium temperature.
Using the fact that $<E> = 0$ at $\b = 2\pi$ and combining with the lower bound
\rr{finalentropylower}, we have:
\bbb
{{\pi c\ll{\rm tot}}\over{12}} \leq
\lno \s\rba\ll{\b = 2\pi} \leq {{\pi c\ll{\rm tot}}\over{12}} + {\rm ln}\lrdd
{{48}\over{48 - c\ll{\rm tot}}} \rrdd \ .
\een{finalentropyupper}

\subsection{Discussion}

We have proven a bound on the number of marginal operators
for a general unitary CFT with discrete spectrum, no relevant
operators other than the identity,
and central charge in a certain range, $c\ll{\rm L} + c\ll{\rm R} < 48$.
We have also proven upper and lower bounds
for the thermodynamic entropy at inverse temperature $2\pi$, under the same
assumptions.  Ultimately, we would like
to improve the bounds, both by extracting more refined information, and
by proving similar bounds under weakened hypotheses.

In terms of
more refined information, it would be good to be able to bound
the number of scalar marginal operators alone, without
including spin-1 operators of weight 2 into the same count. It seems
possible that this could be done by using refined medium-temperature
equations that consider the partition function as a function of
$\t$ and $\bar{\tau}$ instead of ${\rm Im}(\t)$ alone, and using
the stronger condition $\lno (\bar{\tau}\pp\ll{\bar{\tau}})\uu {p\ll 1}
(\tau\pp\ll\tau)\uu{p\ll 2} Z[\t]\rba\ll{\t = i} = 0$ for
$p\ll 1 + p\ll 2$ odd.

As far as weakening the assumptions is concerned,
it would be good to be able to
bound the thermodynamic entropy and microscopic state degeneracies for
arbitrarily high central charge.  Also, the condition that there be no
relevant operators is a strong one, and it is certainly desirable to
derive limits on state degeneracies without this
assumption.  Attempts to weaken either of these two conditions will
necessarily meet some of the difficulties enumerated in the third
section.  New ideas may be required to circumvent those.

In one particular circumstance, the restriction to theories without relevant
operators seems particularly natural.  Two-dimensional CFT with
extended supersymmetry and integrally quantized $U(1)$ charges play
an important role as theories representing string propagation
in spaces with unbroken spacetime SUSY, for instance on
Calabi-Yau threefolds.

These CFT may be a particularly tractable special class
in which the number of marginal operators may be bounded.
First of
all, these theories posess a large chiral algebra -- at least ${\cal N} = 2$
superconformal symmetry, together with spectral flow generators \cite{elliptic2}.
They have low central charge,
circumventing the fine-structure problem discussed earlier.
As for
the continuum problem, conformal sigma models on Calabi-Yau spaces admit non-thermal
boundary conditions under which the partition function can be
evaluated, which project out the contribution from
the near-continuum of operators that may be present with
dimension close to 0.  Each non-thermal boundary conditions for
the partition function nonetheless
has simple modular transformation properties.  Among the
possible boundary conditions for the partition function
are those describing the elliptic genus
\cite{elliptic1, elliptic2}.
Interesting work has been done recently
\cite{mooreetal, manschot, eguchi1, eguchi2}
on the derivation of
general consistency conditions for elliptic genera.  Perhaps
the medium-temperature techniques discussed here combined with
results such as \cite{mooreetal, manschot}
may yield
useful information about the still mostly-uncharted landscape of
Calabi-Yau manifolds.  We hope the present note will provide clues for further
progress in that direction and others.

\begin{center}{\bf Acknowledgments}\end{center}

We are very grateful to many people for valuable discussions and comments on the draft,
including Michael Douglas, Matthias Gaberdiel, Zohar Komargodski,
Hirosi Ooguri, Matthew Sudano and Erik Tonni.  We would also like to thank the Physics Department at Virginia Tech
for hospitality while this work was in progress.
S. H. is supported by the World Premier International Research
Center Initiative (WPI Initiative), MEXT, Japan; and by a Grant-in-Aid
for Scientific Research (22740153) from the Japan Society for Promotion of
Science (JSPS).  C. S.-C. is supported by a fellowship from the Swiss National
Science Foundation (SNF).



\end{document}